\documentclass[pre,amsmath,preprint]{revtex4}
\usepackage{amsfonts}
\usepackage{graphicx}
\usepackage{epsfig}
\begin{document}

\draft
\bibliographystyle{revtex4}

\titlepage
\title{IMPROVED TIME DOMAIN METHOD TO MEASURE NEAR-FIELD DISTRIBUTION OF BURIED-OBJECT}

\author{Li-Ming Si}
\email{10701100@bit.edu.cn}
 \affiliation{Department of Electronic Engineering, School of Information Science and Technology,
 Beijing Institute of Technology, Beijing 100081, People's Republic of China}

\begin{abstract}
The paper analyses the ground antenna echo using microwave frequency
detector and high speed sampling technology and a new method
detecting buried objects in time domain near-field is presented. The
method detecting particular reflection echo frequency of microwave
pulse via digital signal processing is to reduce the false alarm
rate. Simulation results show that this method has advantages of
easy identification and high precision.

\end{abstract}
 \maketitle

\baselineskip=20pt
\section{\label{sec:level1}INTRODUCTION}

Recently, the accuracy of buried-objects detection system is
increasing important which decided personal safety in mine affected
areas.  Few technology is employed in the real world to detect land
mines because of high false alarm rate [1]. Since mine detection
requires low false alarm rate, improving the precision of detection
system should be studied. Many research works have been reported in
the control of false alarm rate. Angular correlation function [2]
and a statistical approach method [3] are employed to detect buried
objects.

Traditional method of detecting buried objects is to analysis the
reflection echo amplitude, which is a analog signal detecting method
and difficult in identifying the existence of several reflection
echo in the lossy medium. A novel digital method detecting
predesignated reflection echo frequency is presented to detect
buried objects.

\section{\label{sec:level2}THEORY AND MEASURE PRINCIPLE}

With the development of finite time difference domain (FDTD) theory,
it has been used to study time domain near-field measurement. FDTD
as a numerical method to solve Maxwell's equations was introduced by
Yee in 1966 [4], which divided both space and time into discrete
grids. The electromagnetic parameters can be figured out in
combination with the boundary conditions. The curl equations that
are used in the FDTD algorithm are
\begin{eqnarray}
  \bigtriangledown\times\mathbf{E}=-\mu\frac{\partial{\mathbf{H}}}{\partial{t}}
  \qquad
  \bigtriangledown\times\mathbf{H}=\varepsilon\frac{\partial{\mathbf{E}}}{\partial{t}}+\sigma\mathbf{E}
\end{eqnarray}
where $\mu$ is permeability, $\varepsilon$ is permittivity and
$\sigma$ is electric conductivity.

 For a two dimensions free space time domain near-field
measurements system, we assume that the dielectric media is
nonmagnetic, \textit{i.e.} $\mu=\mu_{0}$, the $\bf{H}$-field and
$\bf{E}$-field can be written as [5,6]
\begin{equation}
H_x(i,j,t+1)=H_x(i,j,t)-\frac{dt}{\mu_{0}dy}[E_z(i,j,t)-E_z(i,j-1,t)]
\end{equation}

\begin{equation}
H_y(i,j,t+1)=H_y(i,j,t)-\frac{dt}{\mu_{0}dx}[E_z(i,j,t)-E_z(i-1,j,t)]
\end{equation}

\begin{equation}
E_z(i,j,t+1)=\frac{\varepsilon_\infty}{\varepsilon_\infty+\chi_0(i,j)}
E_z(i,j,t)+\frac{1}{\varepsilon_\infty+\chi_0(i,j)}\sum_{m=0}^{t-1}E_z(i,j,t-m)\Delta\chi_m(i,j)+f(H_x,H_y)
\end{equation}
and
\begin{eqnarray}
f(H_x,H_y)&=&\frac{dt}{[\varepsilon_\infty+\chi_0(i,j)]\varepsilon_0dx}[H_y(i+1,j,t)-H_y(i,j,t)] \nonumber \\
          & & -\frac{dt}{[\varepsilon_\infty+\chi_0(i,j)]\varepsilon_0dy}[H_x(i,j+1,t)-H_x(i,j,t)]
\end{eqnarray}
where $\varepsilon_s$ is dielectric's static permittivity,
$\varepsilon_\infty$ is dielectric's optical permittivity, $\mu_0$
is dielectric's permeability, $t_0$ is dielectric's relaxation time
,and
\begin{equation}
{\chi_0(i,j) \choose
\Delta\chi_m(i,j)}=(\varepsilon_s-\varepsilon_\infty){1-exp(\frac{-dt}{t_0})
\choose exp(\frac{-mdt}{t_0})[1-exp(\frac{-dt}{t_0})]^2}
\end{equation} is the susceptibility function.

A periodic microwave pulse has particular characters such as high
peak value and low average value in near field, so it is difficult
to detect the buried objects using amplitude detecting in the lossy
medium. Therefor, to measure high peak value rapidly is the key
point for mine detection system.

we use microwave frequency detector and high-speed sampling
technology in receiver to realize the high efficiency detection. The
component of detection system is shown in Figure 1.
\begin{figure}[!hbt]
\centering
 \epsfig{figure=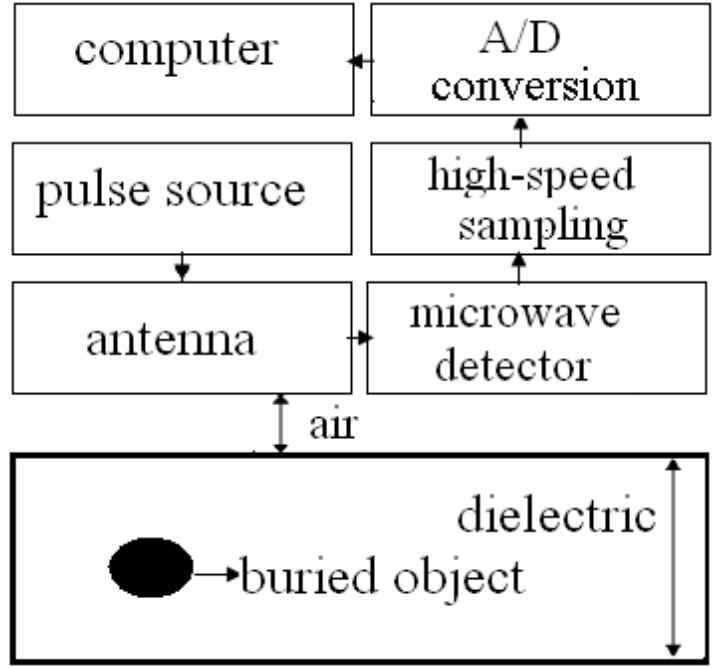}
\renewcommand{\figurename}{Figure}
\caption{Schematic diagram of detection system}
\end{figure}

Microwave pulse was produced by microwave signal source and launched
by transmitting antenna. The antenna can receive reflected pulse
envelop at the same time.

Through microwave detector, we received time domain waveform  of
reflect pulse envelop. This effect is demonstrated in Figure 2.
$\Delta$$t$ is the microwave pulse width and $T_0$ is the period of
microwave pulse.
\begin{figure}[!hbt]
\centering \epsfig{figure=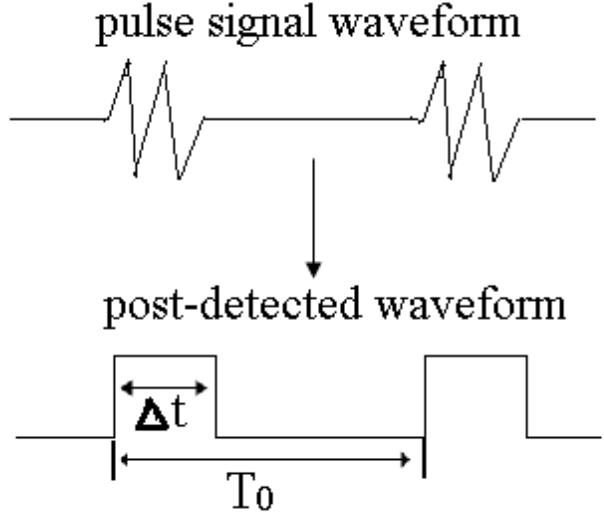}
\renewcommand{\figurename}{Figure}
\caption{Schematic diagram of
pulse signal detector}
\end{figure}

The high-speed sampling is key technology for pulse information
during digital signal processing. High-speed $\bf{A}$/$\bf{D}$
conversion circuit will records high level When the reflection echo
of pulse is detected and records low level during other time.

Finally, completed digital form time domain waveforms can be shown
in the computer.

\section{\label{sec:level3}SIMULATION AND RESULTS}

An iron ball ($\varepsilon_r=12$) with diameter is $15cm$ buried in
a $120cm\times$$120cm\times$$45cm$ cubical vessel filled with dry
sand ($\varepsilon_s=2.5$). As is shown in the Figure 3. There are
$15cm$ between air-dielectric interface with spherical center,
$10cm$ between $P_1$ with air-dielectric interface, $10cm$ between
$P_1$ with $P_2$ and $10cm$ between $P_2$ with $P_3$.
\begin{figure}[!hbt]
\centering \epsfig{figure=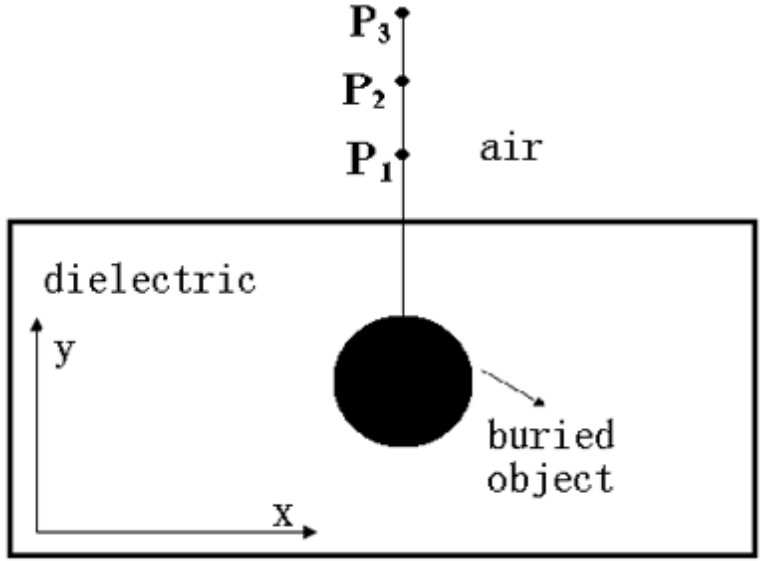}
\renewcommand{\figurename}{Figure}
\caption{The near-field detecting configuration}
\end{figure}

It is well known that FDTD utilizes the Yee cell for calculation at
nodes of the finite-difference lattice. We set 420 grids in the
$X$-direction and 100 grids in the $Y$-direction with a grid spacing
$dx=dy=0.5cm$ in the Yee cell.

We have chosen a microwave signal source with microwave pulse width
$\Delta$$t$$=$$1\mu$$s$ and pulse period $T_0=1ms$ as microwave
pulse producer. A $X$-band microtrip antenna is employed to both
transmit and receive microwave pulse. There is a microwave frequency
detector to detect the $10GHz$ microwave at the receiving end. Then,
we can transform analog signals into digital signals through
high-speed sampling and $\bf{A}$/$\bf{D}$ conversion. Finally, the
completed time domain near-field information of refection echo is
shown in the computer monitor.

We put the microtrip antenna at $P_1$, $P_2$ and $P_3$ to detect
buried object respectively.

The analog signal simulation result is described in Figure 4 and
digital signal simulation result in Figure 5.
\begin{figure}[!hbt]
\centering
{\label{a}\epsfig{figure=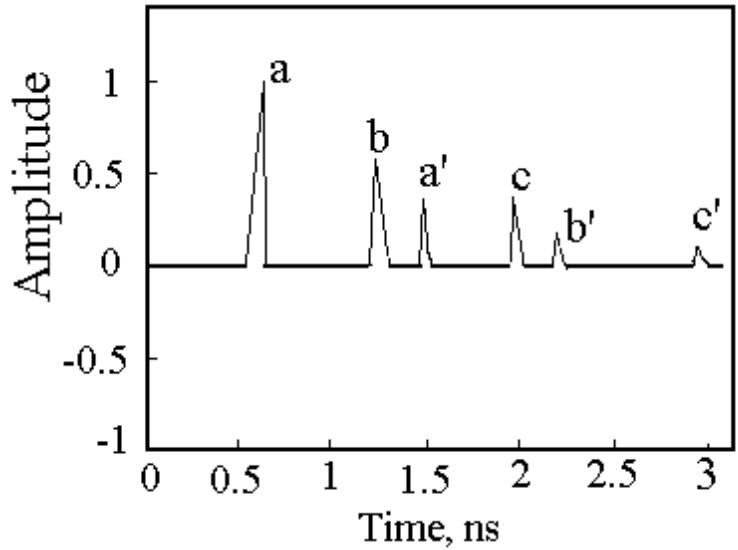}}
\renewcommand{\figurename}{Figure}
\caption{Analog signal simulation results}
\end{figure}

\begin{figure}[!hbt]
\centering
{\label{b}\epsfig{figure=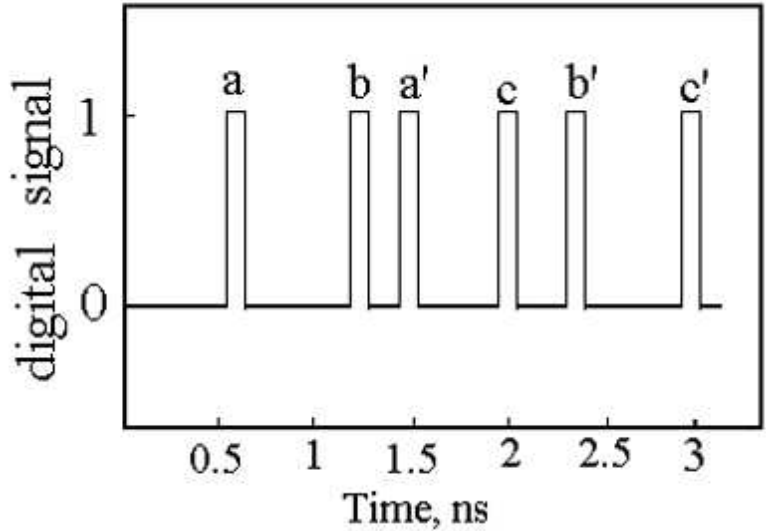}}
\renewcommand{\figurename}{Figure}
\caption{Digital signal simulation results}
\end{figure}

The peaks $a$ and $a^{'}$ are the detecting results at $P_1$, $b$
and $b^{'}$ are the detecting results at $P_2$ and $c$ and $c^{'}$
are the detecting results at $P_3$.

\section{\label{sec:level4}CONCLUSION}

The paper introduces a frequency detection method using microwave
frequency detector and high-speed sampling technology to detect the
buried-objects in time domain near-field. From the simulation
results we can see that the method can get more clear and accurate
reflection echo information from the buried-objects than traditional
amplitude detecting method. At the same time, the low false alarm
rate detecting method would be a new idea to detect remote objects.


\newpage
{\bf{Figure Captions}}

Figure 1 Schematic diagram of detection system

Figure 2 Schematic diagram of pulse signal detector

Figure 3 The near-field detecting configuration

Figure 4 Analog signal simulation results

Figure 5 Digital signal simulation results
 \noindent
\end{document}